\documentclass[aps,prb,twocolumn,superscriptaddress,showpacs,amsmath,amssymb,longbibliography]{revtex4-2}
\usepackage[english]{babel}
\usepackage{amsmath}
\usepackage{bm}
\usepackage{graphicx,bbm} 
\usepackage{times}
\usepackage{epsfig} 
\usepackage{soul}
\usepackage[utf8]{inputenc}
\usepackage[colorlinks,linkcolor=blue,citecolor=blue,urlcolor=blue]{hyperref}
\usepackage{color}
\usepackage{latexsym}
\usepackage{ulem}
\definecolor{nred} {RGB}{224,0,0}
\definecolor{nblue} {RGB}{28,130,185}

\begin{document} 
\title{Diffusion in the Anderson model in higher dimensions}
\author{P. Prelov\v{s}ek}
\affiliation{Jo\v zef Stefan Institute, SI-1000 Ljubljana, Slovenia}
\affiliation{Faculty of Mathematics and Physics, University of Ljubljana, SI-1000 Ljubljana, Slovenia}
\author{J. Herbrych}
\affiliation{Department of Theoretical Physics, Faculty of Fundamental Problems of Technology, Wroc\l aw University of Science and Technology, 50-370 Wroc\l aw, Poland}

\date{\today}
\begin{abstract}
We present an extended microcanonical Lanczos method (MCLM) for a direct evaluation of the diffusion constant and its frequency dependence within the disordered Anderson model of noninteracting particles. The method allows to study systems beyond $10^6$ sites of hypercubic lattices in $ d = 3- 7$ dimensions. Below the transition to localization, where we confirm dynamical scaling behavour, of interest is a wide region of incoherent diffusion, similar to percolating phenomena and to interacting many-body localized systems. 
\end{abstract}

\maketitle

\noindent {\it Introduction.} The metal-insulator (MI) transition in disordered systems of noninteracting fermions is established and theoretically a well understood phenomenon since the fundamental work of Anderson \cite{anderson58}, the scaling theory of localization \cite{wegner76,abrahams79}, and numerous analytical and numerical studies captured in several reviews \cite{kramer93,markos06,evers08,suntajs21}. Since the MI transition exists only in lattices of higher dimensions $d \geq3$, the focus of numerical efforts was in the analysis of the critical behavior, primarily of the localization length $\xi$ and its critical exponent $\nu$, which is by now even quantitatively well established within the standard Anderson model in $d=3$ \cite{mackinnon81,mackinnon83,kramer93,slevin18}, but also for higher $d\leq 6$ \cite{rodriguez10,garcia07,pietracaprina16,mard17,tarquini17}. The transport properties of the disordered system were first approached via the sensitivity to boundary conditions \cite{edwards72,thouless74} resulting in an important concept of Thouless energy and time scale in finite (also interacting many-body) systems. On the other hand, numerical studies and explicit results of intrinsic properties as the optical conductivity $\sigma(\omega)$ \cite{weisse04,weisse06}, with related d.c. conductivity $\sigma_{0}$ \cite{mackinnon81,economou85,weisse04} and diffusion coefficient $D_0$ \cite{prelovsek78,prelovsek79,ohtsuki97,markos06,sierant20} are surprisingly sparse, also due to the lack of powerful numerical methods.

In the past decade interest in disordered models revived in connection with the challenging phenomenon of the many-body localization (MBL) \cite{basko06,oganesyan07,znidaric08,berkelbach10,pal10,barisic10,huse14,luitz15,lev15,serbyn15} which predicts the MI transition also in $d=1$ system, i.e., in the Anderson disordered model with interaction between fermions (or equivalently in the anisotropic Hesenberg spin chain). The connection between Anderson and MBL models has been recently reinvestigated \cite{sierant20} in a wide range of disorder and $d=3,5$, also in terms of the characteristic Thouless time $\tau_\mathrm{Th} \propto L^2/D_0$ (where $L$ is the system length) and related Thouless energy $E_{\mathrm{Th}}=2\pi/\tau_\mathrm{Th}$ \cite{bertrand16,schiulaz19,suntajs20,sonner20}.

In this Letter we present a numerical method for an efficient calculation of the dynamical diffusion coefficient $D(\omega)$, and in particular its d.c. value $D_0$, within the Anderson model of $d$-dimensional disordered lattice. The method is the extension of the microcanonical Lanczos method (MCLM) \cite{long03,prelovsek11} employed already within numerous studies of (mostly high-temperature $T \gg 0$) transport in MBL models \cite{karahalios09,barisic10,mierzejewski16,prelovsek17}. Here, we use the method for $T \to 0$ diffusion of non-interacting (NI) particles and adapt it for very high frequency resolution $\delta \omega$ and for hypercubic lattices beyond $N = L^d \sim 10^6$ sites. This allows us to scan $D_0$ as well as $D(\omega)$ from the weak-scattering regime up to localization transition at $W=W_c$ for dimensions $d=3-7$. Results reveal in all $d$ three distinct regimes: a) the weak-scattering region for small $W < W^*$, b) the critical regime $W \lesssim W_c$ following the scaling behavior, and c) very wide intermediate regime, in particular for $d>3$, with small and incoherent $D(\omega)$ with effective mean free-path $\lambda <1$, reminiscent of a percolative diffusion. The latter transport has similarities, but also differences, to the (sub)diffusive regime in MBL systems. On the localized side of the MI transition we employ the method to study the dynamical imbalance $C(\omega)$ and related d.c. value $C_0$, the quantity experimentally studied in MBL cold-atom systems \cite{schreiber15}, including the case of NI disordered systems \cite{bordia16}, but also closely related to experiments on classical waves in continuous disordered systems \cite{hildebrand14}.

We consider the standard Anderson model \cite{anderson58} of NI fermions on a $d$-dimensional hipercubic lattice with the on-site quenched disorder, 
\begin{equation}
H = - t\sum_{\langle ij \rangle} \left( c^\dagger_{j} c_i + \mathrm{H.c.}\right)
+ \sum_i \epsilon_i c^\dagger_{i} c_i\,,
\label{mod} 
\end{equation}
where the hopping is between nearest-neighbor (n.n.) lattice sites and random local energies are assumed to have uniform distribution $-W/2 < \epsilon_i <W/2$. We will use theoretical units $\hbar=1$, $t$ as a unit of energy, and lattice spacing $a_0=1$. We focus only on the physics in the middle of the spectrum, i.e., at energies ${\cal E} \sim 0$, where also values for critical disorder strength $W_c$ are well established, i.e., $W_c/t \sim 16.5 $ \cite{rodriguez10,slevin18}, for $d=3$ up to $W_c/t \sim 83 $ \cite{tarquini17} for $d=6$.

\noindent {\it Numerical approach to diffusion.} 
The dynamical conductivity, being isotropic in the hypercubic lattice, can be expressed in a system of NI fermions with of the Kubo-Greenwood formula \cite{thouless74},
\begin{equation}
\sigma(\omega) = \frac{\pi e_0^2}{N \omega} \sum_{n,m} [f_n-f_m]
|\langle \varphi_n| J |\varphi_m \rangle|^2 
\delta(\omega - E_m+E_n)\,,
\label{kg}
\end{equation}
where the current operator $J = t \sum_{i} ( i c^\dagger_{i+1_x} c_i + \mathrm{H.c.}) $ is taken for convenience in one ($x$) direction, assuming also periodic boundary conditions in all directions. $E_n$, $|\varphi_n\rangle$ are fermion eigenenergies and eigenfunctions, respectively, and $f_n = 1/[\mathrm{e}^{(E_n-{\cal E})/(k_B T)}+1]$ is the state occupation for given Fermi energy ${\cal E}$ and temperature $T$. For a hypercube $N=L^d$ is the number of sites. At $T \to 0$ the d.c conductivity $\sigma_0 = \sigma (\omega \to 0)$ depends only on eigenstates with $E_{n,m} \sim {\cal E}$ and it is convenient to express it with the d.c. diffusion coefficient $D_0$ as $\sigma_0 = e_0^2 {\cal N}_F D_0$, where ${\cal N}_F$ is the density of states at ${\cal E}$. Since we are interested in the low frequencies $\omega \lesssim t$ (smaller than an effective band-width) Eq.~(\ref{kg}) yields an expression for $D(\omega)$,
\begin{equation}
D(\omega) = \frac{\pi}{N} \sum_{m} |\langle \varphi_n| J |\varphi_m \rangle|^2 
\delta(\omega - E_m+E_n)\,,
\label{dom}
\end{equation}
provided that $E_n \sim {\cal E}$ and that the resulting $D(\omega)$ (in the macroscopic limit $L \to \infty$) is a self-averaging quantity, i.e., is independent of chosen $|\varphi_n\rangle$.

Whereas Eq.~(\ref{dom}) in a finite system apparently requires a full exact diagonalization (ED) of the model (\ref{mod}), and in particular the knowledge of the eigenfunction $|\varphi_n \rangle$, we use at this point the idea of MCLM method \cite{long03,prelovsek11} and replace $|\varphi_n \rangle$ with the single microcanonical state $|\Psi_{\cal E} \rangle $ with the energy ${\cal E}$. The latter is within MCLM obtained via the Lanczos-type approach using the operator $V=(H-{\cal E})^2$. Performing $M_L \gg 1$ Lanczos iterations the result should converge well for the lowest eigenstate of $V$. Since in the present application we have in mind Hilbert spaces with typically $N_{st} \gtrsim 10^6$ states, such a Lanczos procedure is not expected to converge to an eigenstate, but rather to a state with very small energy dispersion $\sigma^2_{\cal E} = \langle \Psi_{\cal E} |V| \Psi_{\cal E} \rangle$. By performing Lanczos procedure twice and also extracting only the lowest eigenfunction of $V$, the storage of the emerging three-diagonal matrix is needed without final ED of $M_L\times M_L$ matrix. This allows us to use large $M_L \sim 10^5$ necessary to get high resolution $\sigma_{\cal E}/t < 10^{-4}$. The second step is then the evaluation of the correlation function, Eq.~(\ref{dom}), as resolvent,
\begin{equation}
D(\omega) = \frac{1}{N} \mathrm{Im} \langle \Psi_{\cal E}| 
J \frac{ 1}{\omega - i \eta + {\cal E} - H} J |\Psi_{\cal E} \rangle\,,
\label{dom1}
\end{equation}
The latter is evaluated with the Lanczos procedure for $H$, starting with $J |\Psi_{\cal E} \rangle$ as an initial vector, which after $M_L$ iterations gives Eq.~(\ref{dom1}) in terms of continued fractions, evaluated finally using an appropriate damping $\eta \gtrsim \delta \omega$. Within such MCLM procedure the frequency resolution is directly connected to $M_L$ as $\delta \omega \leq \Delta E/M_L$ where $\Delta E$ is the energy span of $H$ within chosen finite-size system. For given $M_L$ we typically also get $\sigma_{\cal E} < \delta \omega$. 

For the study of transport and dynamical correlations in the Anderson NI model, where $N_{st} = N$ it is essential to reach besides large Hilbert spaces with $N > 10^6$, also high frequency resolution with typically $\delta \omega/t < 10^{-4}$, representing long-time dynamics up to $\tau \sim 1/\delta \omega > 10^4/t$. Within presented MCLM this is achieved by optimizing the choice of $N$ and $M_L$ whereby the limitations are given mostly by CPU time $\propto N M_L$, while memory requirement in considered models is determined by $N z$ where $z =2d$ is the connectivity of $H$, i.e., the number of n.n. in the lattice. In the following we present results for the Anderson model typically with $N \gtrsim 10^6$ sites and $M_L \sim 10^5$ iterations which for modest $W$ leads to $\delta \omega/t \sim 10^{-4} $. We note that such numerical approach to $D_0$ is more convenient than, so far mostly used, time-evolution of the wavepacket spread \cite{prelovsek79,ohtsuki97,markos06,sierant20}, since the latter requires open boundary conditions and hardly can reach times $ \tau > 10^3/t$.

\begin{figure}[tb]
\includegraphics[width=1.0\columnwidth]{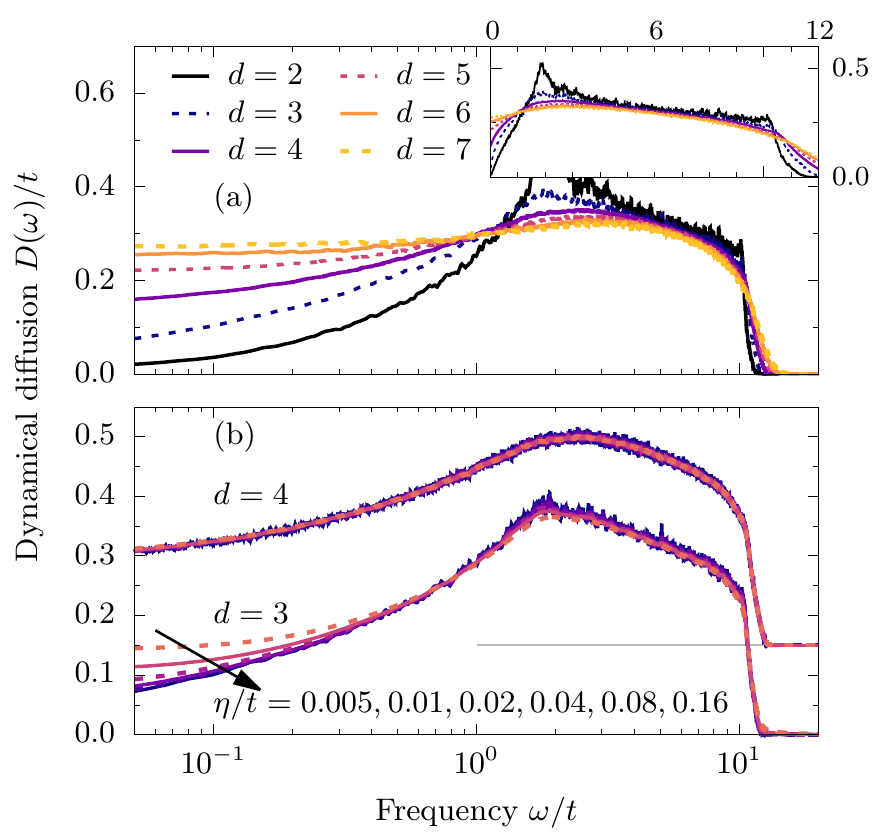}
\caption{ Dynamical diffusion $D(\omega)$ in the Anderson model at the intermediate disorder $W/t=20$, a) for hypercubic lattices $d = 2- 7$, and b) for $d=3,4$ showing the influence of the damping $\eta$. }
\label{fig1}
\end{figure}

\noindent{\it Diffusion coefficient: results.} 
Before turning to the d.c. transport let us consider some general features of dynamical $D(\omega)$. We note that our diffusion $D(\omega)/t$ is dimensionless since $D \propto a_0^2/\tau_0$ and $\tau_0=\hbar/t$ and we have chosen $\hbar=a_0=1$. In Fig.~\ref{fig1}a we present typical spectra for intermediate disorder $W/t =20$, calculated for dimensions $d=2-7$. The case is chosen so that for $d= 3$ it is $W \gtrsim W_c$, for $d>3$ disorder is subcritical $W<W_c$, while in $d=2$ all states are localized. It is evident that high-frequency dynamics $D(\omega/t > 1)$ is essentially $d$-independent, with spectra extending to $\omega \propto W$ [note that in this regime $D(\omega)$ does not reflect directly $\sigma(\omega)$]. The localized cases , i.e., $d=2,3$, typically reveal large spectral fluctuations and require sampling over disorder realizations $M_s \gg 1$. On the other hand, $D(\omega)$ at $\omega/t <1$ and in particular $\omega \to 0$ are clearly $d$-dependent, and as shown in Fig.~\ref{fig1}b the resolution and choice of small $\eta/t \ll 1$ is crucial to reproduce small $D_0\ll t$ or even localized regime with $D_0 =0$ as is the case for $d=3$ at $W/t =20$.

\begin{figure}[tb]
\includegraphics[width=1.0\columnwidth]{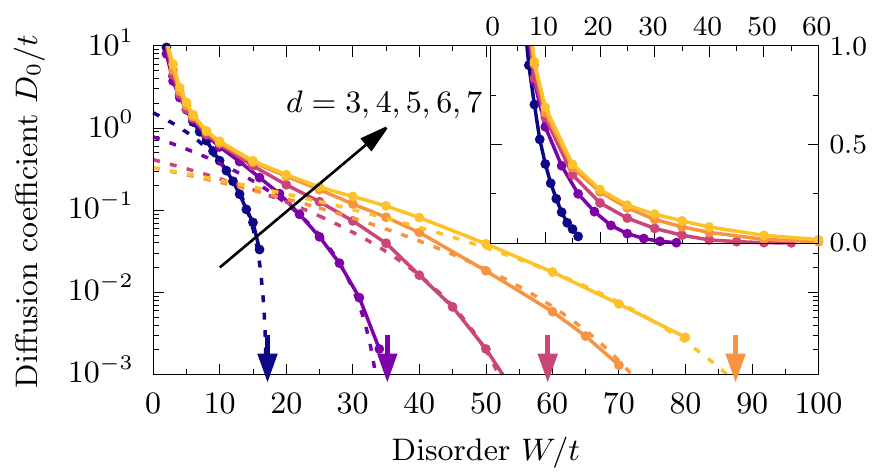}
\caption{ Diffusion coefficient $D_0$ vs. disorder strength $W$ within the Anderson model in hypecubic lattices with $d = 3 - 7$ (in the inset in the normal scale for $D_0/t<1$) for $D_0$, in the vicinity of critical regime also fitted with the scaling form $D_0 \propto (W_c-W)^s$ (see text for details). }
\label{fig2}
\end{figure}

The central quantity of this Letter is the d.c. diffusion coefficient $D_0$ in the middle of the band ${\cal E}=0$ and for $d=3-7$. This is calculated via MCLM on isotropic lattices with $N = L^d$ sites using in the evaluation of resolvent, Eq.~(\ref{dom1}), at $\omega=0$ the damping $\eta \gtrsim \delta \omega$. The result is $\eta$-sensitive only in the cases with strong $D(\omega)$ dependence, which is actually the case at $W \sim W_c$ in $d=3,4$. We present here results for $N \gtrsim 10^6$, i.e., $L=100, 36, 16, 12, 8$ for $d=3-7$, respectively. Considered quantity $D(\omega)$ is expected to be self-averaging (unlike the conductance \cite{slevin01,slevin03,mierzejewski21}) for $L \to \infty$. In spite of large systems studied, we still observe at $W \lesssim W_c$ sample-to-sample fluctuations of $D_0$, so we employ also a modest sample averaging with $M_s=10-30$.

Results for $D_0$ vs. $W$ are presented in Fig.~\ref{fig2}. It is evident that the method allows to follow $D_0$ for more than three decades, where its lower bound is mostly determined by reachable $\delta \omega$ at chosen $N$. It is characteristic that we reach lowest $D_0/t \sim 10^{-3}$ for $d=5$ due to less singular $D(\omega)$ (discussed later-on), while for $d=6,7$ small $D_0$ might be already limited by finite-size effects. Still, results in Fig.~\ref{fig2} clearly reveal three different regimes of diffusion: 

\noindent a) Weak-scattering regime, for all $d$ - typically at $W<W^*\sim 10 t$ - we confirm $D_0 = c_d /W^2$, where $c_d \propto {\cal N}_F $. Since considered disorders $W\geq 2$ already smear most details of density of states ${\cal N}({\cal E})$, one could expect ${\cal N}_F t\propto 1/\sqrt{2\pi z}$. However, results on Fig.~\ref{fig2} seem to indicate even weaker $d$ dependence. Here we note, that (as standard) defined $D(\omega)$, Eq.~(\ref{dom}) refers to a propagation in only one ($x$) direction, so it should be quite $d$-independent in the regime $W<W^*$.
 
\noindent b) Wide intermediate regime, particularly well pronounced for higher dimensions $d \geq 4$, where the diffusion is incoherent, i.e., $D_0/t < 1$ in all $d$ at $W > W^*$. Since $D_0 = \bar v_x \lambda_x$, where particle effective velocity (in one direction) $\bar v_x \sim t$ and $\lambda_x$ is the corresponding transport mean free path, this regime implies $\lambda_x < 1$. It is rather surprising that such transport persists in such a wide range of $W < W_c$. It even indicates on some universal form $D_0 \propto \mathrm{exp}(-c W)$ for $d \geq 5$, as pointed out recently \cite{sierant20}, having the similarity to the variation of d.c. conductivity $\sigma_0$ \cite{barisic10,prelovsek17} and the inverse Thouless time \cite{suntajs20} in the MBL prototype model (see also discussion later-on). 

\noindent c) The critical regime $ W \lesssim W_c$ is characterized in Fig.~\ref{fig2} as the drop from quasi-linear $\mathrm{ln} (D_0/t) $ vs. $W$ dependence, whereby $W_c$ is increasing with $d$. Close to the MI transition results can be well captured with $D_0 \propto (W_c-W)^s$ and $s=(d-2)\nu$ from the scaling theory \cite{wegner76,abrahams79} and critical disorder values $W_c/t \sim 16.5, 35, 59, 87, 107$ and localization-length exponents $\nu \sim 1.57, 1.1, 0.96, 0.84. 0.72 $ for $d =3 - 7$, respectively, well consistent with focused numerical studies of the Anderson transition \cite{mackinnon83,rodriguez10,garcia07,pietracaprina16,mard17,tarquini17,slevin18}. Also, our results appear to be consistent with decreasing $\nu \to 0.5$ for $d \to \infty$ \cite{garcia07,mard17}.

\noindent{\it $D(\omega)$: critical regime.} 
Although in the weak-scattering regime $W<W^*$ $D(\omega)$ is essentially Lorentzian with relaxation rate $1/\tau \propto W^2$, in the intermediate regime $W^*<W< W_c$ spectra are broad and quite featureless, with nearly constant low-frequency value $D(\omega <1) \sim D_0$, as shown in Fig.~\ref{fig1}. Frequency dependence becomes nontrivial in the critical regime where we can test it with the scaling form $\sigma(\omega) = \xi^{2-d} F(\xi/L_\omega)$ \cite{shapiro81}, where $L_\omega \propto [D(\omega)/\omega]^{1/2}$ is the characteristic length scale (at given $\omega$) for density correlation. This suggests the relation
\begin{equation}
D(\omega) = w^s F\bigl(w ^{-\nu} \sqrt{\frac{\omega}{D(\omega)}} \bigr)\,,
\label{crit}
\end{equation}
where $w=(W_c-W)/W_c$, and for the scaling function we assume a simple form $F(x)= A+ Bx^{d-2}$ \cite{shapiro81}, satisfying both limits: a) $w > 0, \omega \to 0$ with $D_0 = A w^s$, discussed already in connection with Fig.~\ref{fig2}, b) $w \to 0, \omega >0$, where the relation, Eq.~(\ref{crit}), yields $D(\omega) \sim B ~\omega^p$ with $p=(d-2)/d$. 

\begin{figure}[tb]
\includegraphics[width=1.0\columnwidth]{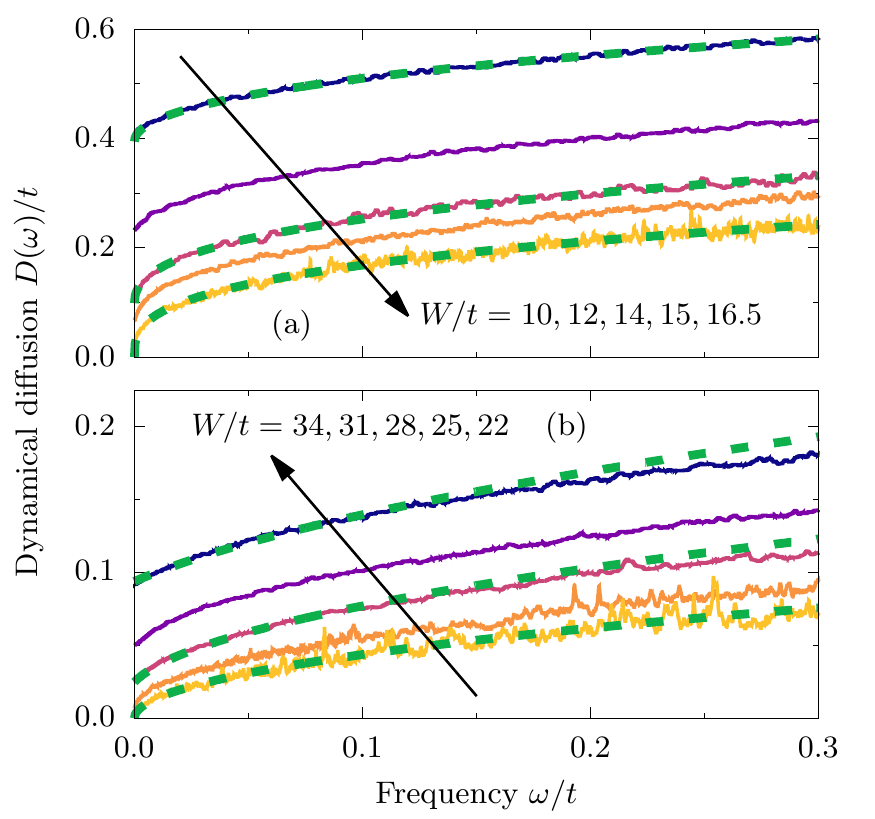}
\caption{ Dynamical diffusion response $D(\omega)$ in the vicinity of the Anderson transition $W \lesssim W_c$ compared to the scaling form $D(\omega) = w^s F( w^{-\nu} L_\omega)$ for a) $d=3$, and b) $d=4$ Anderson model (see text for details). }
\label{fig3}
\end{figure}

In Fig.~\ref{fig3} we present our numerical result for $D(\omega)$ for several values $W$ in the critical regime $W \lesssim W_c$ for $d=3$ and $d=4$. Results restricted to the window $\omega \ll 1$ are shown along with the solution of Eq.~(\ref{crit}) with fixed $A,B$. For $d=3$ our results in Fig.~\ref{fig3}a are well consistent with anomalous $D(\omega) \propto \omega^{1/3}$ at critical $w=0$, turning into $D(\omega) \sim D_0 + \alpha \sqrt{\omega}$ at $w>0$ \cite{shapiro81}. We note that steep $\omega$ dependence at $w \gtrsim 0$ is also preventing us from reaching small values of $D_0$ in $d=3$, as compared to $d \geq 4$ data, as evident in Fig.~\ref{fig2}. In contrast, results for $d=4$ in Fig.~\ref{fig3}b follow expected $D(\omega) \propto \sqrt{\omega}$ at $w \sim 0$ as well as $D(\omega) \sim D_0 + \gamma \omega$ for $w>0$. We also find that for $d>4$ at the MI transition $D \propto \omega^p$ where $ p =1- 2/d \to 1$ with increasing $d > 3$.

\noindent{\it Imbalance.} 
On the insulating side of MI transition, $W>W_c$, we can also apply our MCLM method to evaluate dynamical quantities. Since in this regime $D_0=0$, of interest at $ \omega \to 0$ are time-dependent density correlations $C(t) \propto \langle \rho_{\bf q} (t)
\rho_{\bf q} \rangle$ and their Fourier transform 
\begin{equation}
C(\omega) = \frac{1}{N} \mathrm{Im} \langle \Psi_{\cal E}| \rho_{\bf q} 
\frac{ 1}{\omega-i \eta + {\cal E} - H}
\rho_{\bf q} |\Psi_{\cal E} \rangle\,,
\label{com}
\end{equation}
where $\rho_{\bf q}= \sum_i \mathrm{e}^{i {\bf q}\cdot{\bf R}_i} n_i$ is the density modulation operator. In connection with theory of MBL systems \cite{luitz16,mierzejewski16,prelovsek17}, as well as related experiments on cold-fermion systems \cite{schreiber15,bordia16}, the quantity of interest is the imbalance which probes ${\bf q}/\pi= \mathbf{1}_d$ response (with $\mathbf{1}_d$ as a d-dimensional unity). In the localized regime one expects a singular response with $C(\omega)= C_0 \delta(\omega) + C_{reg}(\omega)$, where $C_0$ is the imbalance stiffness. We note that $C_0$ has been directly measured in cold-atom chains \cite{bordia16}, but is closely related also to analogous infinite-range intensity correlations investigated in $d=3$ disordered classical-wave systems \cite{hildebrand14,shapiro99}.

Here, we concentrate on $C_0$ which reflects the localization of Anderson wavefunctions, and in particular should - in the critical regime - behave as the inverse localization length $C_0 \propto 1/\xi \propto w^{\nu}$. Such quantity should be self-averaging even in the random system, in contrast to, e.g., local density correlation $C_{ii}(\omega)$ (analogous to inverse participation ratio). The MCLM results discussed below reveal substantial sample-to-sample fluctuations of $C_0$, since by choosing small $\sigma_{\cal E}$ we effectively get $C_0$ averaged only over $N_{eff}= \sigma_{\cal E} N$ Anderson localized states, generating significant statistical error in the localized regime. 

\begin{figure}[tb]
\includegraphics[width=1.0\columnwidth]{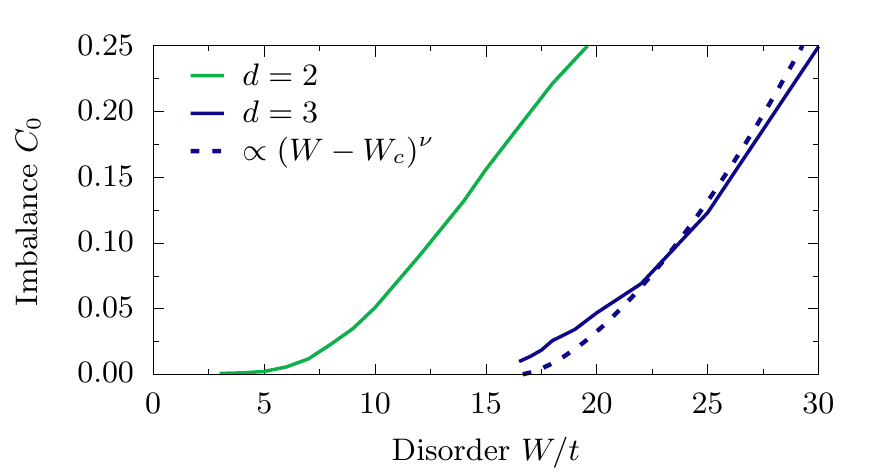}
\caption{ Imbalance stiffness $C_0$ vs. $W$ for Anderson model in $d=2,3$ dimensions. Results for $d=3$ are fitted to critical behavior $C_0\propto (W-W_c)^\nu$.}
\label{fig4}
\end{figure}

In Fig.~\ref{fig4} we present results for $C_0$ vs. $W$ for $d=2,3$. Since results reveal larger sample-to-sample fluctuations, here we take smaller $N \sim 3.10^5$, but larger $M_s \sim 100$ and presented $C_0$ are average values. It should be noted that $C_0$ are by definition normalized for NI particles, $\int d\omega C(\omega) =1 $, so in the extreme localization limit $C_0=1$. Although in $d=3$ results can be well described by the critical behavior of the localization length, i.e., $C_0 \propto w^{\nu}$ with $\nu = 1.57 $, in $d=2$ the variation of $C_0$ vs. $W$ remains finite $C_0$ at $W>0$, but still with a sharp crossover at $W^*/t \sim 7$ with the onset of stronger localization at $W>W^*$.

Let us finally in more detail comment on similarities as well as differences to physics of the MBL systems: 

\noindent a) {\it Incoherent diffusion: percolation.} From Fig.~\ref{fig2} it is evident that beyond $W>W^* \sim 10 t$ there is is wide span of $W$, particularly pronounced for $d \geq 4$, with the incoherent diffusion characterized by mean-free path $\lambda_x < 1$. We note that the marginal $W^* \sim B$ can be related to an effective bandwidth, scaling roughly as $B \sim 2 \sqrt{z} t$. In order to capture qualitatively the incoherent regime $W^* < W <W_c$, we can employ simple concept of propagation through resonant states, which allows to make contact with transport emerging in MBL systems, due to interaction between localized NI Anderson states \cite{prelovsek21}. At $W^* \gg t$ the diffusion in the Anderson NI model can appear through the resonance between n.n. sites. Probability for these sites to satisfy the resonance $|\epsilon_i-\epsilon_j| \lesssim 2 t$ is $P_1 \sim 2 t/W \ll1 $. Taking into account the connectivity $z=2d$ and requiring the overall probability $P_1 \sim 1$, one can reach marginal $W_1^* \sim 2z t < W_c$, at least in $ d \geq 4$. For $W>W_1^*$ diffusion in higher $d >3$, hopping to further neighbors via intermediate sites becomes relevant. E.g., for next n.n. hop between $i,j$ via intermediate site $k$, we get $\tilde t_{ij} \sim t^2/(\epsilon_i-\epsilon_k)$. The effective total hopping probability $P_2 \sim z^2 p_2$ is then obtained via perturbation theory (where lower-resonances $\epsilon_i-\epsilon_k<2t$ are omitted),
\begin{equation}
P_2 = z^2 \frac{2 t_2}{W}, \qquad t_2 \sim \frac{t^2}{W} 
\int_{2t}^W \frac{d \Delta}{\Delta } = \frac{ t^2}{W} \mathrm{ln} \frac{W}{2t}\,.
\qquad 
\end{equation}
Requiring $P_2 \sim 1$ yields critical \mbox{$W_2^* \propto zt[2\ln(W/2t)^{1/2}]$}. One can continue such estimates taking into account further neighbors and higher resonances with effective hopping $t_n = (t^n/W^{n-1}) \ln^{n-1}(W/2t_{n-1})$. Such procedure leads to known estimate for the critical disorder $W_c \propto 2 z t \ln(W_c/2t) $ \cite{ziman69,thouless74}.

Although the above derivation is just a rough counterpart of the original arguments \cite{anderson58,thouless74} for the convergence of perturbation expansion in the localized regime $W>W_c$, our aim here is to connect the phenomenon of the incoherent diffusion to transport in MBL systems. In the latter systems, the prototype being the $d=1$ disordered chain of interacting fermions \cite{basko06,oganesyan07,znidaric08,berkelbach10} the interaction allows the hopping between Anderson states \cite{prelovsek18}, typically localized on next n.n. and further neighbors. Such process has analogy to percolation problem in high-$d$ lattice \cite{prelovsek21}. Although from above arguments we cannot establish an analytical dependence of $D_0(W)$, it is evident from Fig.~{\ref{fig2} that in high $d \geq 5$ it can be reasonably represented as $D_0 \propto \exp(-c W/t)$ \cite{sierant20}, although with much smaller $c \ll 1$ compared to MBL models where $c \sim 1$ \cite{barisic10,prelovsek17}. 

\noindent b) {\it $D(\omega)$: subdiffusion.} Strictly, the phenomenon of subdiffusion requires $D_0=0$ and $D(\omega) \propto \omega^p$ with $p<1$. In MBL models the existence of such transport (in the ergodic regime) is still controversial. On the other hand, in the NI Anderson model this is the case (only) at the critical point, where $p = (d-2)/d$, while for $W<W_c$ this is just a transient feature (e.g. in time) \cite{sierant20} since $D_0>0$. Again, $D(\omega)$ resembles MBL systems more for $d \gg 3$ since $p \to 1$, which is the situation of dynamical conductivity $\sigma(\omega)$ at the presumed transition into the localized phase \cite{barisic10,prelovsek17}.

\noindent{{\it Summary}.} We introduced a numerical method which allows the study of dynamical correlation functions in nontrivial models of NI particles, reaching larger sizes as well as high-frequency resolution. The method has promising potential also for application in similar problems requiring both large Hilbert spaces and high frequency resolution as ,e.g., MBL and (nearly)-integrable models. We focused here on the dynamical diffusion $D(\omega)$ in the Anderson model in hypercubic $d \geq 3$ lattices, where also the MI transition exists. Our d.c. and dynamical results are in the critical regime $W \sim W_c$ well consistent with the scaling theory of localization. On the other hand, we find a broad regime of incoherent diffusion which has similarities as well differences with the challenging problem of many-body localization.

\noindent{\it Acknowledgments}
P.P. acknowledges the support of the project N1-0088 of the Slovenian Research Agency. J.H. acknowledges support by the Polish National Agency of Academic Exchange (NAWA) under contract PPN/PPO/2018/1/00035.

\bibliography{manuandd}
\end{document}